\documentclass[review,number,sort&compress]{elsarticle}
\usepackage{lineno}

\usepackage{graphicx}
\usepackage{amssymb}
\usepackage{color}
\usepackage{url}
\usepackage{ulem}
\usepackage[figuresright]{rotating}

\definecolor{orange}{rgb}{1,0.5,0}

\journal{Nuclear Instruments and Methods A}

\begin{document}

\begin{frontmatter}


\title{A First Comparison of the responses of a $^{4}$He-based fast-neutron  
detector and a NE-213 liquid-scintillator reference detector} 


\author[arktis]{R.~Jebali\fnref{fn2}}
\author[lund,ess]{J.~Scherzinger}
\author[glasgow]{J.R.M.~Annand}
\author[arktis]{R.~Chandra}
\author[arktis]{G.~Davatz}
\author[lund,ess]{K.G.~Fissum\corref{cor1}}
\ead{kevin.fissum@nuclear.lu.se}
\author[arktis]{H.~Friederich}
\author[arktis]{U.~Gendotti}
\author[ess,midswe]{R.~Hall-Wilton}
\author[lund]{E.~H\aa kansson}
\author[ess]{K.~Kanaki}
\author[m4]{M.~Lundin}
\author[arktis]{D.~Murer}
\author[ess,m4]{B.~Nilsson}
\author[m4]{A.~Rosborg}
\author[m4,sweflo]{H.~Svensson}

\address[arktis]{Arktis Radiation Detectors Limited, 8045 Z\"{u}rich, Switzerland}
\address[lund]{Division of Nuclear Physics, Lund University, SE-221 00 Lund, Sweden}
\address[glasgow]{University of Glasgow, Glasgow G12 8QQ, Scotland, UK}
\address[ess]{Detector Group, European Spallation Source ESS AB, SE-221 00 Lund, Sweden}
\address[m4]{MAX IV Laboratory, Lund University, SE-221 00 Lund, Sweden}
\address[midswe]{Mid-Sweden University, SE-851 70 Sundsvall, Sweden}
\address[sweflo]{Sweflo Engineering, SE-275 63 Blentarp, Sweden}

\cortext[cor1]{Corresponding author. Telephone:  +46 46 222 9677; Fax:  +46 46 222 4709}
\fntext[fn2]{present address: University of Glasgow, Glasgow G12 8QQ, Scotland, UK}

\begin{abstract}
A first comparison has been made between the pulse-shape discrimination 
characteristics of a novel $^{4}$He-based pressurized scintillation detector 
and a NE-213 liquid-scintillator reference detector using an Am/Be mixed-field 
neutron and gamma-ray source and a high-resolution scintillation-pulse digitizer. 
In particular, the capabilities of the two fast neutron detectors to discriminate 
between neutrons and gamma-rays were investigated. The NE-213 liquid-scintillator 
reference cell produced a wide range of scintillation-light yields in response 
to the gamma-ray field of the source. In stark contrast, due to the size and 
pressure of the $^{4}$He gas volume, the $^{4}$He-based detector registered a 
maximum scintillation-light yield of 750~keV$_{ee}$ to the same gamma-ray field. 
Pulse-shape discrimination for particles with scintillation-light yields of more 
than 750~keV$_{ee}$ was excellent in the case of the $^{4}$He-based detector. 
Above 750~keV$_{ee}$ its signal was unambiguously neutron, enabling particle
identification based entirely upon the amount of scintillation light produced.
\end{abstract}


\end{frontmatter}

\section{Introduction}
\label{section:introduction}

Fast neutrons are important both as probes of matter and as diagnostic
tools~\cite{walker82,homeland03,FNDA06,FNDA11,ESS09,chandra10,lyons11,
chandra12,peerani12,FNASS13,islam13,lewis13,tomanin14,lewis14}. In the case 
that information about the energy and emission time of a neutron is available, 
conclusions about its origin can be drawn. The timing precision required 
to obtain this information may only be provided by neutron detectors that 
are fast, providing signals with short risetimes. Today, organic liquid 
scintillators are the detectors-of-choice for fast neutrons. Drawbacks 
associated with these scintillators are their toxicity, reactive nature, 
and sensitivity to a broad range of gamma-ray energies.

Scintillators are substances which emit light when subjected to ionizing
radiation. The characteristic time constant associated with the light emitted
is a function of the properties of the scintillator in question. Certain
scintillators respond to different types of ionizing radiation differently;
that is, the time constant of the emitted light is different depending upon
the density of ionization produced by the incident radiation. Normally,
there are several components with different time constants. The relative
intensity of these components affects the effective integrated time 
constant.  By carefully analyzing the behavior 
of the scintillation light as a function of time, one can determine the 
incident particle type. This procedure is called pulse-shape discrimination
(PSD). PSD is often used to distinguish between different types of uncharged
particles, namely gamma-rays and neutrons. In scintillators with good PSD 
properties, incident gamma-rays interact primarily with the atomic electrons 
of the scintillator, producing close to minimum-ionizing electrons which give
a fast (decay times of some 10s of ns) flash of light. On the other hand, 
incident neutrons interact primarily with the hydrogen in liquid scintillators 
and $^{4}$He nuclei in noble-gas scintillators via scattering, transferring 
some of their energy. For hydrogen, this energy transfer can be 100\%, while
for $^{4}$He, the energy transfer is at best 64\%.  The resulting flashes 
of light arising from the much denser ionization produced by the relatively 
large energy loss of the 
recoiling protons and alpha particles have longer decay times (100s to 1000s 
of ns). PSD and thus incident particle identification may be performed by 
recording the time dependence of the scintillation pulse form and comparing 
the fast and slow components.

\section{$^{4}$He as a scintillation medium for fast neutron detection}
\label{section:he4}

The development of both liquid and gaseous $^{4}$He based scintillators  
for fast-neutron detection has been 
reported~\cite{mckinsey03,chandra10,chandra12,lewis13,lewis14}. $^{4}$He, 
like most noble gases, is a good scintillator. It has an ultra-violet light 
yield comparable to the intrinsic non-Tl doped light yield of NaI 
crystals~\cite{aprile06,birks64,knoll89,dolgosheim69}. Neutron interactions 
lead to $^{4}$He recoils, where energy is deposited very locally within the gas. 
Gamma-ray interactions lead to recoiling electrons, which deposit only tens of 
keV per centimeter of trajectory. This difference in deposition density and 
therefore ionization density is believed to ultimately enable the PSD capability. 
PSD properties may be degraded significantly if the geometry and size of the 
detector results in a smearing of the transit times of scintillation photons 
comparible to the scintillation decay times. Good PSD also requires good 
scintillation efficiency; that is, a sufficient number of scintillation photons 
to define the time dependence of the pulse accurately, and low noise in the 
pulse-processing electronics.

With only two electrons per atom, $^{4}$He has a very low charge density, 
thereby significantly limiting its sensitivity to gamma-rays. This is useful 
for fast-neutron detection, where insufficient gamma-ray rejection is often the 
factor which constrains the desired performance. The following physical effects 
contribute positively to the gamma-ray rejection performance of pressurized 
$^{4}$He gas:
\begin{enumerate}
\item {\it Low gamma-ray interaction probability.} Due to the low electron 
density of $^{4}$He, gamma-ray interaction probabilities are two orders of 
magnitude lower than neutron interaction probabilities.
\item {\it Low energy deposition.} Depending on the chosen geometry (i.e.~a 
tube with radius of a few cm), the amount of energy the gamma-rays can deposit 
in the detector volume is limited. This is because the corresponding Compton 
or pair electrons cannot transfer much energy to the gas before striking a 
detector wall.
\item {\it Similar scintillation-light yield for gamma-rays and neutrons.} 
The scintillation light 
production in organic liquid scintillators is highly velocity dependent. Thus, 
for the same amount of deposited energy, gamma-ray interactions produce more 
scintillation light than neutron interactions. In contrast, the 
scintillation-light yield for gamma-rays and neutrons is similar in noble-gas 
scintillators~\cite{birks64} such as $^{4}$He. $^{4}$He is commonly called 
a linear scintillator.
\item {\it~PSD.} 1--3 above together with the fast and slow components of the 
$^{4}$He scintillation signals lead to excellent PSD and thus excellent separation of neutrons and gamma-rays.
\end{enumerate}
\noindent
The purpose of this project was to compare the neutron/gamma discrimination
obtained using the Arktis $^{4}$He-based neutron-diagnostic tool (NDT) to that 
obtained using a reference liquid-scintillator cell filled with the organic 
liquid scintillator NE-213~\cite{ne213}. 
\noindent

\section{Apparatus}
\label{section:apparatus}

\subsection{Am/Be source}
\label{subsection:ambe_source}

The detector characterizations reported on in this paper were carried out 
using a nominal 18.5 GBq $^{241}$Am/$^{9}$Be (Am/Be) 
source~\cite{hightechsoltd} which emitted (1.106$\pm$0.015)~$\times$~10$^6$ 
neutrons per second nearly isotropically~\cite{natphyslab}. The source is a 
mixture of americium oxide and beryllium metal contained in an X.3 capsule,
which is a stainless-steel cylinder 31~mm (height) $\times$ 22.4~mm 
(diameter)~\cite{x.3}. $^{241}$Am has a half-life of 432.2 years and decays 
via alpha emission (5 discrete energies with an average value of about 5.5 MeV) 
to $^{237}$Np. The dominant energy of the gamma-rays associated with the decay 
of the intermediate excited states in $^{237}$Np is $\sim$60 keV.
A 3~mm thick Pb sheet was used to complement the stainless steel X.3 capsule to
attenuate these 60~keV gamma-rays. The half-value layer for Pb for 60~keV 
gamma-rays is 0.12~mm, while for 1~MeV gamma-rays, it is 8~mm.  Neutrons are 
produced when the emitted alpha particles undergo a nuclear 
reaction with $^{9}$Be resulting in $^{12}$C and a free neutron. The resulting 
neutron distribution has a maximum energy of about 11~MeV~\cite{lorch73}, while
approximately 25\% of the neutrons have an energy of less than 
1~MeV~\cite{vijaya73}. The de-excitation of the $^{12}$C results in a ~4.44 MeV 
gamma-ray about 55\% of the time~\cite{vijaya73,mowlavi04,liu07}. This 
gamma-ray is too energetic to be absorbed by the stainless steel of the X.3
capsule. Thus the radiation field from the Am/Be is a combination of 
high-energy gamma-rays and fast neutrons. Both the gamma-ray and fast-neutron 
dose rates at a distance of 1~m from the source in the unshielded X.3 capsule 
were measured using a Thermo Scientific Corporation FHT 752 dosimetric neutron 
detector~\cite{thermo}. They were both determined to be 11~$\mu$Sv/hr for a 
total unshielded dose rate of 22~$\mu$Sv/hr, in exact agreement with the data 
sheet from the supplier.

\subsection{Arktis pressurized $^{4}$He gas Neutron Diagnostic Tool (NDT)}
\label{subsection:ts8_ndt}

\begin{figure}
\begin{center}
\resizebox{0.40\textwidth}{!}{\includegraphics{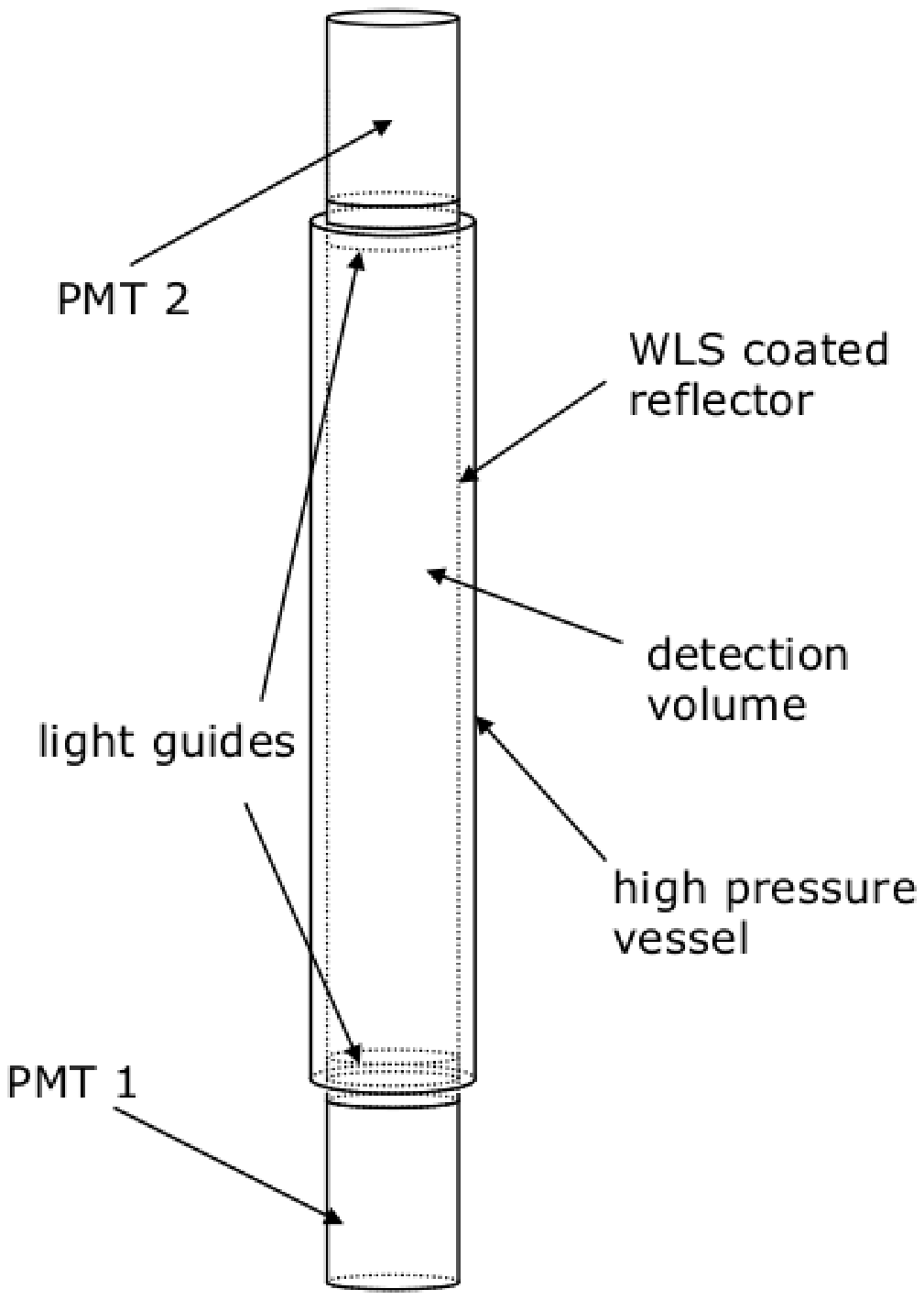}}
\resizebox{0.58\textwidth}{!}{\includegraphics{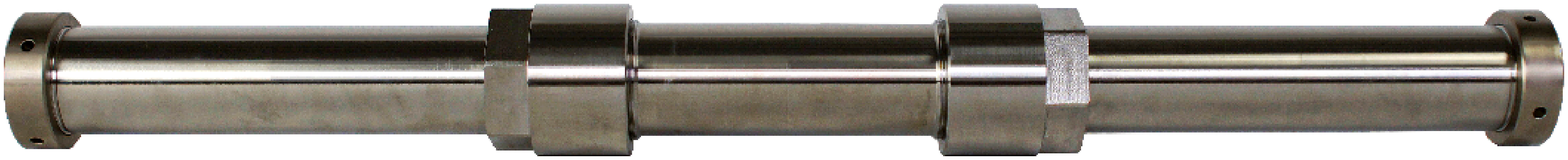}}
\caption{\label{figure:ref_detectorNDT}
Schematic (left) and photograph (right) of the pressurized $^{4}$He gas
fast-neutron detector. The outer diameter was 5.08~cm (2$"$) and the active 
length was 19.5~cm.
}
\end{center}
\end{figure}
\noindent

The version of the Arktis pressurized gas $^{4}$He fast NDT used for these 
measurements is shown in Fig.~\ref{figure:ref_detectorNDT}. It was of cylindrical 
geometry with an outer diameter of 5.08~cm (2$"$) and a 19.5~cm active length.
The detector walls were made of stainless steel. The interior surface of the 
stainless-steel cylinder was coated with a PTFE-based diffuse reflector~\cite{ptfe} 
which was itself coated with an organic phosphor that converted the wavelength 
of the scintillation light from 80~nm to 430~nm. As $^{4}$He is transparent to 
its own light, almost no signal loss due to reabsorption occurs~\cite{mck}. 
The predominant mechanism for signal loss was due to multiple reflections inside 
the detector. Optical windows capable of withstanding the 120~bar operating 
pressure were employed. The scintillation signals were read out at both ends of 
the active volume by Hamamatsu R580~\cite{hama_r580} photomultiplier tubes (PMTs).

\subsection{NE-213 reference detector}
\label{subsection:ne213_reference_detector}

\begin{figure}
\begin{center}
\resizebox{0.70\textwidth}{!}{\includegraphics{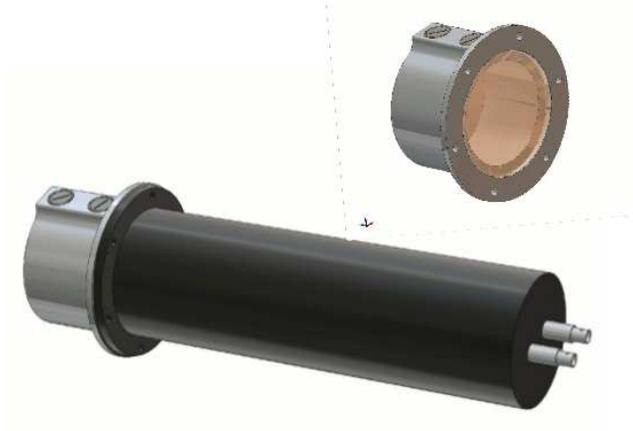}}
\caption{\label{figure:ref_detector}
The NE-213 reference detector. Top: a detail of the cylinder ``cup".  The 
screws on top allow for the filling or draining of the liquid cylinder. 
A borosilicate-glass window (light brown) serves as the optical boundary. 
See text for details. Bottom: The black cylinder to the right is the 
magnetically shielded 3~inch ET Enterprises 9821KB photomultiplier-tube 
assembly. The gray cylinder to the left is the ``cup". (For interpretation 
of the references to color in this figure caption, the reader is referred 
to the web version of this article.)
}
\end{center}
\end{figure}

The NE-213 reference detector is shown in Fig.~\ref{figure:ref_detector}. The
core of the NE-213 reference detector was a 3~mm thick cylindrical aluminum 
cell with an inner depth of 62~mm and an inner diameter of 94~mm. The inside of
the cell was painted with EJ-520~\cite{ej520} titanium dioxide reflective 
paint, which can withstand the xylene solvent of the liquid scintillator. The 
aluminum cell was sealed using a 5~mm thick borosilicate glass 
window~\cite{borosilicate} glued to the aluminum cell using the highly 
temperature and chemical resistant Araldite 2000$+$~\cite{araldite}. The 2 
penetrations into the cell which allowed for filling were sealed with M-8 
threaded aluminum plugs with 20~mm diameter heads and 14~mm diameter Viton 
O-rings~\cite{viton}. Nitrogen gas was bubbled through the NE-213 liquid 
scintillator for 24~hours prior to filling the cell. The assembled cell was 
then filled with the nitrogen-flushed NE-213 using a nitrogen gas transfer 
system. 

After filling, the borosilicate glass window of the cell was coupled to a 
cylindrical PMMA~\cite{pmma} lightguide with a depth of 57~mm and a diameter 
of 72.5~mm. PMMA is an acrylic which transmits light down to 300~nm in 
wavelength~\cite{eljen_pmma}. The cylindrical surface of this lightguide was 
painted with water-soluble EJ-510~\cite{ej510} reflective paint. The lightguide 
was then pressure-coupled via springs to a magnetically shielded 3~inch ET 
Enterprises 9821KB PMT assembly~\cite{et_9821kb}. As our goal was to produce 
a stable detector which provided reproducible results, no optical-coupling 
grease was used. 

\subsection{Arktis WaveDREAM-B16 digitizer}
\label{subsection:arktis_digitizer}

\noindent

\begin{figure}
\begin{center}
\resizebox{0.8\textwidth}{!}{\includegraphics{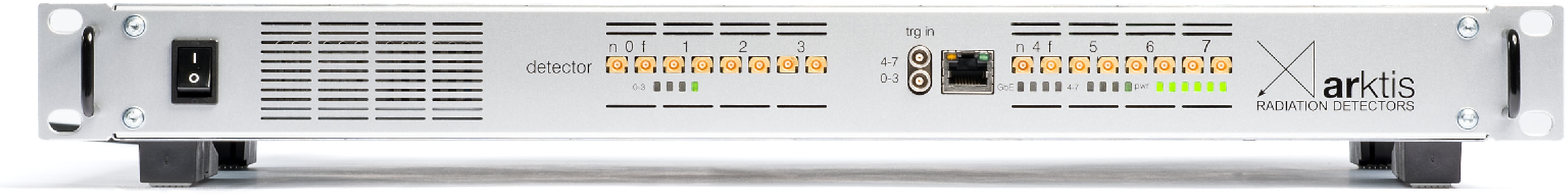}}
\resizebox{0.75\textwidth}{!}{\includegraphics{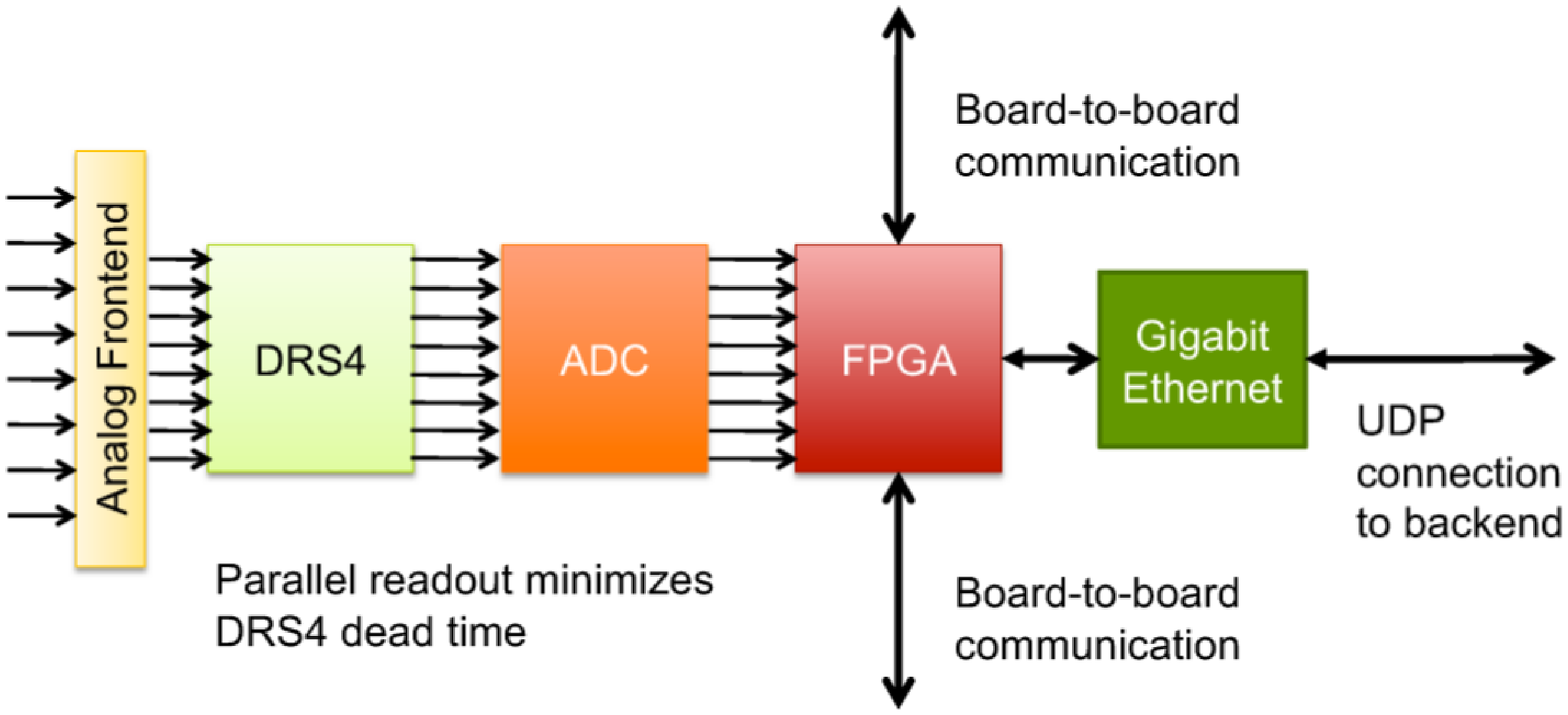}}
\resizebox{0.75\textwidth}{!}{\includegraphics{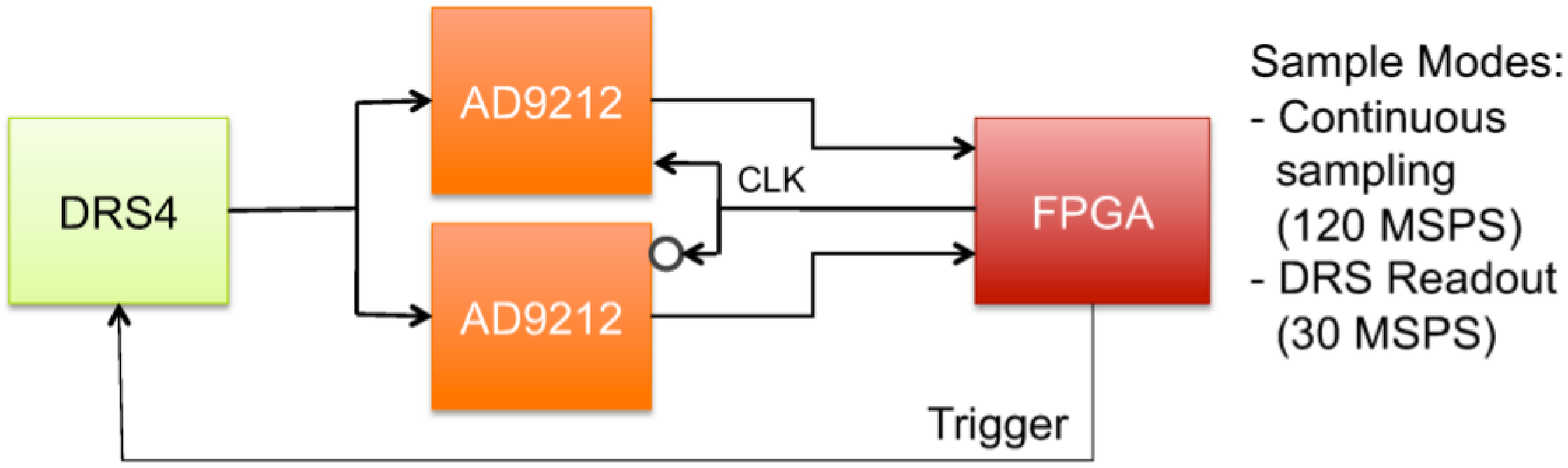}}
\caption{\label{figure:wd16}
The top figure shows the WaveDREAM-B16. There are 16 input 
channels grouped in 2 clusters of 8. Once the selected signals are digitized, 
they are read out via Gigabit ethernet. Triggers can be applied externally or 
generated internally by the software. The middle figure is a schematic overview 
of the readout electronics. Signals are stored in the DRS4 switched-capacitor 
array and read out if user-defined trigger conditions are met. The bottom 
figure presents an overview of the software trigger. If the signal sampled at 
120~MSPS meets the trigger condition, the FPGA reads out the DRS4 at 
1~GSPS. See Ref.~\cite{friederich11} for further details. 
}
\end{center}
\end{figure}

The analog signals from the NDT and NE-213 reference cell were fed directly to 
a WaveDREAM-B16 high-precision digitizer developed by Arktis Radiation 
Detectors~\cite{arktis}~(see Fig.~\ref{figure:wd16}). After being converted to 
a digital signal, the data of interest were stored for analysis. The decision 
for storage was based on a 120~megasample per second (MSPS) signal that was 
continuously read out and fed into a field programmable gate array (FPGA).
If the user-defined trigger condition was met (see below for the trigger
conditions employed in this measurement), the FPGA read out the DRS4 switched 
capacitor array~\cite{ritt} containing the stored waveform at 1~gigasample 
per second (GSPS). Complete details are presented in Ref.~\cite{friederich11}.
In this manner, excellent time resolution (provided by the 1~GSPS sampling and 
nanosecond time stamping also between different detectors) was achieved at 
10-bit resolution over the 3.5 $\mu$s duration of the stored event. The trigger 
conditions were user-configurable. In the case of the NDT, a trigger occurred 
only if the signals from both PMTs mounted on each end of the detector were 
above threshold. This threshold was 35 scintillation photons detected above 
the ambient baseline, and suppressed the dark-current count rate. 
An event was considered to be valid if the second PMT signal came less than 
32~ns after the first, a criterion related to the speed of light, dimensions 
of the $^{4}$He gas volume, and response of the PMTs. This broad timing window 
in the digitizer was inherited from studies of a different system and was 
considered to be ``safe" for the investigations of the much smaller NDT studied 
here. In the case of the NE-213 reference cell which had only one PMT, a 
trigger occurred if the analog signal coming from the PMT was above threshold. 
This threshold was 30 scintillation photons detected above the ambient baseline. 
The digitizer greatly facilitated the offline analysis of the data as it allowed 
for the Arktis NDT PMTs to be synchronized to better than 1~ns. It also allowed 
for offline variation in the detector thresholds and integration gates (see below), 
a clear advantage over standard discriminators and analog-to-digital converters. 

\section{Measurement}
\label{section:measurement}

\subsection{Setup}
\label{subsection:setup}

\begin{figure}
\resizebox{1.00\textwidth}{!}{\includegraphics{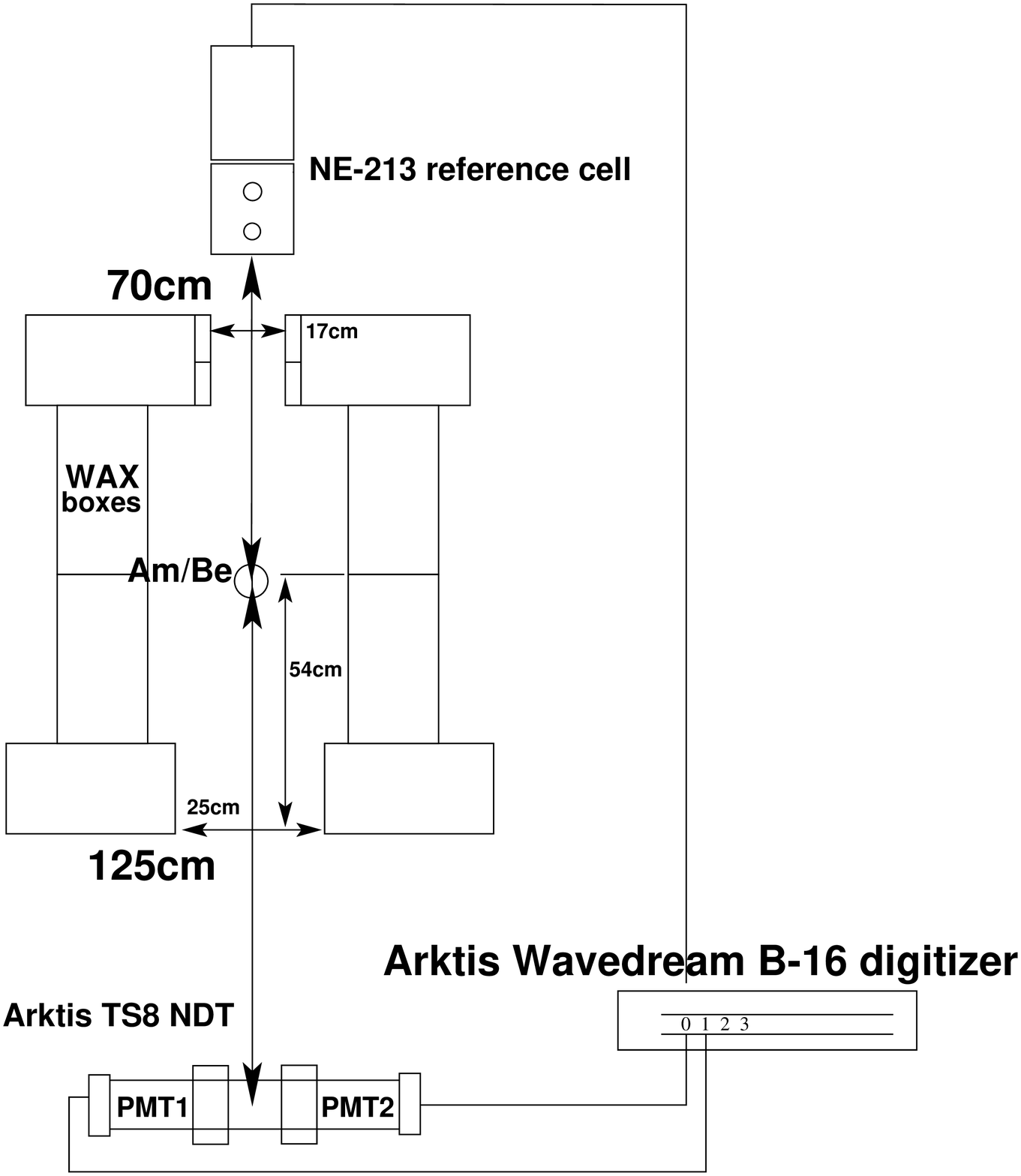}}
\caption{\label{figure:LundSetup_4}
The experiment setup (not to scale). The Am/Be source was placed at the center 
of a borated-wax enclosure. Penetrations through two of the enclosure walls allowed 
for a direct line-of-sight between the source and the detectors. For measurements 
of the gamma-ray distribution at the apertures, the NE-213 reference cell was 
replaced with a stand-alone HPGe detector.
}
 
\end{figure}

The experiment setup is shown in Fig.~\ref{figure:LundSetup_4}.  The Am/Be
source in its transport/storage container was placed at the 
center of a 4-sided enclosure constructed from borated-wax boxes. In the 
so-called ``park" position with the source locked in its transport container 
at the bottom of the borated-wax box enclosure, the total dose rate in the 
room was less than 0.4~$\mu$Sv/hr. When the source was lifted from its container 
and positioned within the 3~mm Pb thick sleeve, the gamma-ray dose rate on the 
outside of the 4-sided borated-wax enclosure was 1~$\mu$Sv/hr.

\begin{figure}
\centering
\resizebox{0.76\textwidth}{!}{\includegraphics{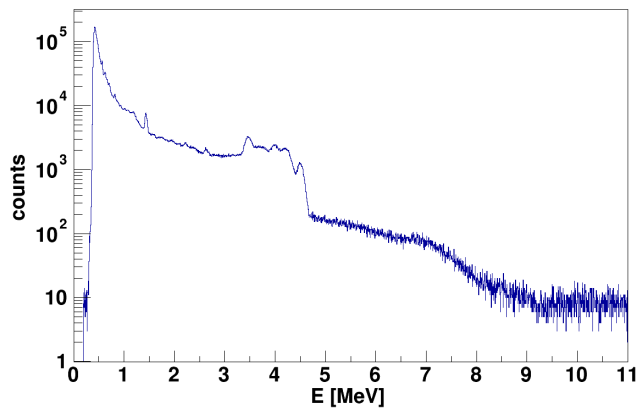}}
\caption{\label{figure:screened_AmBe_gammas}
Energy distribution of gamma-rays at the approximate location of the NDT and
NE-213 reference cell. The prominent structure between 3 and 5~MeV results
from the 4.44~MeV gamma-ray.
}
\end{figure}

Square penetrations through 2 of the opposite walls of the enclosure allowed 
direct line-of-sight between the detectors being irradiated and the source. 
A HPGe gamma-ray detector was positioned in one of the apertures. The distance
to the screened source was $\sim$1~m and line-of-sight was direct. It was used 
to measure the distribution of gamma-rays at the approximate location of the 
NDT and NE-213 reference cell. Fig.~\ref{figure:screened_AmBe_gammas} presents 
this distribution, where the 4.44~MeV gamma-ray peak together with its Compton 
edge, first- and second-escape peaks, and their Compton edges are clearly seen 
between (from the right) 5 and 3~MeV. Between 1.0~MeV and the $\sim$400~keV 
detector threshold, the gamma-ray intensity increased by an order-of-magnitude.

At a distance of 0.54~m from the source, the square aperture for the NE-213 
reference detector was 17 $\times$ 17~cm$^{2}$. The cell was placed at source 
height so that the face of the active volume (recall 
Fig.~\ref{figure:ref_detector}) was 70~cm from the center of the Am/Be source. 
The cylindrical symmetry axis of the detector pointed directly at the center 
of the source.  It was operated at $-$1700~V. The analog signals from the 
single PMT were fed into the digitizer. The digitized waveforms of the signals 
which triggered the acquisition were recorded on an event-by-event basis for 
offline processing. At the same distance from the source, the square aperture 
for the NDT was 25 $\times$ 25~cm$^{2}$.  The NDT was also placed at source 
height. However, its cylindrical symmetry axis was perpendicular to the ray 
pointing to the center of the source. Both of the PMTs (one at each end of 
the detector volume) were operated at $+$1730~V. The analog signals from these 
PMTs were fed into the digitizer. Again, the digitized waveforms of the signals 
which triggered the acquisition were recorded on an event-by-event basis for 
offline processing.

A pulse-shape (PS) analysis of the analog signals coming from both detectors 
was performed using the ``tail-to-total" method. Two integration gates were 
defined, a long gate (LG) and a short gate (SG). Both of the gates opened at 
the same time, 10~ns before the analog signal. The SG was used to integrate only 
the fast components of the analog signal, while the LG was used to integrate 
the entire (both fast and slow components) analog signal. The PS was determined 
from the difference between the scintillation-light yield in the LG and SG 
normalized to the scintillation-light yield in the LG: PS = (LG-SG)/LG. For the 
NDT PMT data, the SG was 100~ns (related to the decay constant of the 
fast-scintillation components of $^{4}$He) and the LG was 3500~ns (the maximum 
gate length the digitizer provided). For the NE-213 reference cell data, the SG 
was 25~ns and the LG was 150~ns. The Arktis NDT gate widths were optimized 
offline. This procedure was greatly facilitated by the digitizer.
 
\subsection{Absolute energy calibration}
\label{subsection:energy_calibration}

Energy-calibration measurements for the NE-213 reference cell were performed 
using $^{60}$Co, $^{137}$Cs, and the Pb-shielded Am/Be sources. $^{60}$Co emits
gamma-rays with energies 1.17~MeV and 1.33~MeV. $^{137}$Cs emits a gamma-ray 
with an energy 0.66~MeV. The de-excitation gamma-ray from the first excited 
state of $^{12}$C has an energy of 4.44~MeV. The locations of the Compton edges
from these gamma-rays were determined using the prescription of Knox and 
Miller~\cite{knox72}. 

For the NDT, Geant4 simulations~\cite{geant4a,geant4b} reproduce the shape of the 
observed pulse-height distributions for neutrons and gamma-rays well. 
The $\alpha$ decay of trace amounts of $^{222}$Rn in the $^{4}$He gas of the 
NDT provides an energy signature similar to the neutron/$^{4}$He scattering 
process, and was used to calibrate the detector.  Neutrons, 
which produce a recoiling $\alpha$ particle, were correlated directly to the 5.5, 
6.0, and 7.7~MeV $\alpha$ lines in $^{222}$Rn~\cite{chandra12}. Gamma-rays produce 
an electron via Compton scattering or an electron and positron via pair production. 
These interactions occur dominantly in the relatively high-Z walls of the NDT. 
Apart from very low energies, most electrons or positrons entering the $^{4}$He
gas volume do not stop in the gas. On traversing the gas, they loose a fairly 
well-defined energy of around 150~keV. This produces a peak in the pulse-height 
distribution which can be used for cross calibration. The energy-loss distribution 
extends out to around 750~keV, which is more or less independent of the incident 
gamma-ray energies above 750~keV.  These energy losses are of course dependent 
upon the size and pressure of the $^{4}$He gas volume. From this, we have 
established that the scintillation-light yield is the same for electrons and 
alpha particles, consistent with $^{4}$He being a linear scintillator. This 
contrasts with NE-213, where the scintillation-light yield depends strongly on 
the velocity and ionization density of the interacting particle.

\section{Results}
\label{section:results}

\begin{figure}
\resizebox{1.00\textwidth}{!}{\includegraphics{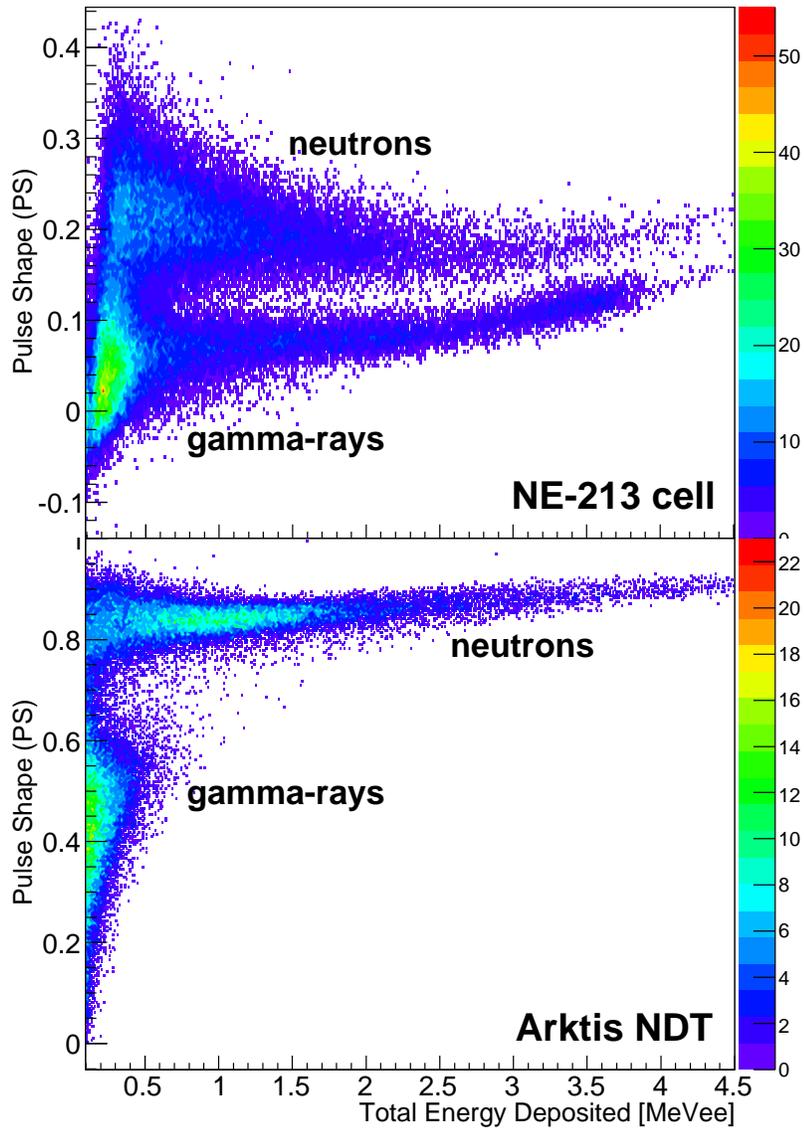}}
\caption{\label{figure:ndt_refcell_psd}
A comparison of the PSD achieved using the NE-213 reference cell (top panel) 
and the $^{4}$He-based NDT (bottom panel) with the Pb-screened Am/Be source. 
For each event, the signal pulse shape PS = (LG-SG)/LG has been plotted against 
the scintillation-light yield produced in the detector for the LG in MeV$_{ee}$.
The distributions corresponding to neutrons and gamma-rays are labeled. 
}
\end{figure}

We stress that the data presented in this section came directly from the digitizer 
and were not optimized via offline software corrections in any way.
Figure~\ref{figure:ndt_refcell_psd} shows a two-dimensional scatterplot comparison 
of the PSD achieved using the NE-213 reference cell and the $^{4}$He-based NDT 
obtained using the Pb-screened Am/Be source. 
Recall that the Pb-screened Am/Be 
source provided a continuous energy spectrum of neutrons up to 11 MeV and a 
of gamma-rays up to 4.44~MeV. The gamma-ray pulse-height response of the NE-213 
reference cell extends to 4~MeV$_{ee}$ as shown in the upper panel. Below 
500~keV$_{ee}$, significant overlap between the neutron and gamma-ray pulse-height 
responses occurred. As shown in the lower panel and in stark contrast, the 
gamma-ray pulse-height response of the Arktis NDT extends only to 750~keV$_{ee}$, 
and clear separation between the neutron and gamma-ray pulse-height responses is
evident down to 100~keV$_{ee}$.

The amount of scintillation light produced in the gaseous $^{4}$He is less than 
that produced in the liquid scintillator for both particle types. In absolute 
terms, the detection efficiency of the NE-213 reference cell will qualitatively 
be higher the Arktis NDT, both for neutrons and gamma-rays. Quantitative 
evaluation of these detection efficiencies and comparisons with Monte Carlo 
calculations will be addressed in a future publication. We note that the lower 
absolute detection efficiency of the Arktis NDT could be advantageous in very 
high intensity radiation fields. The complete lack of a gamma-ray band to higher 
energies in the bottom scatterplot is striking. Relative to the numbers of 
neutrons detected, the gamma-ray discrimination properties of the Arktis NDT 
are clearly superior.  The digitizer clearly proved to be a very effective tool 
for optimizing the PSD.

\begin{figure}
\resizebox{1.00\textwidth}{!}{\includegraphics{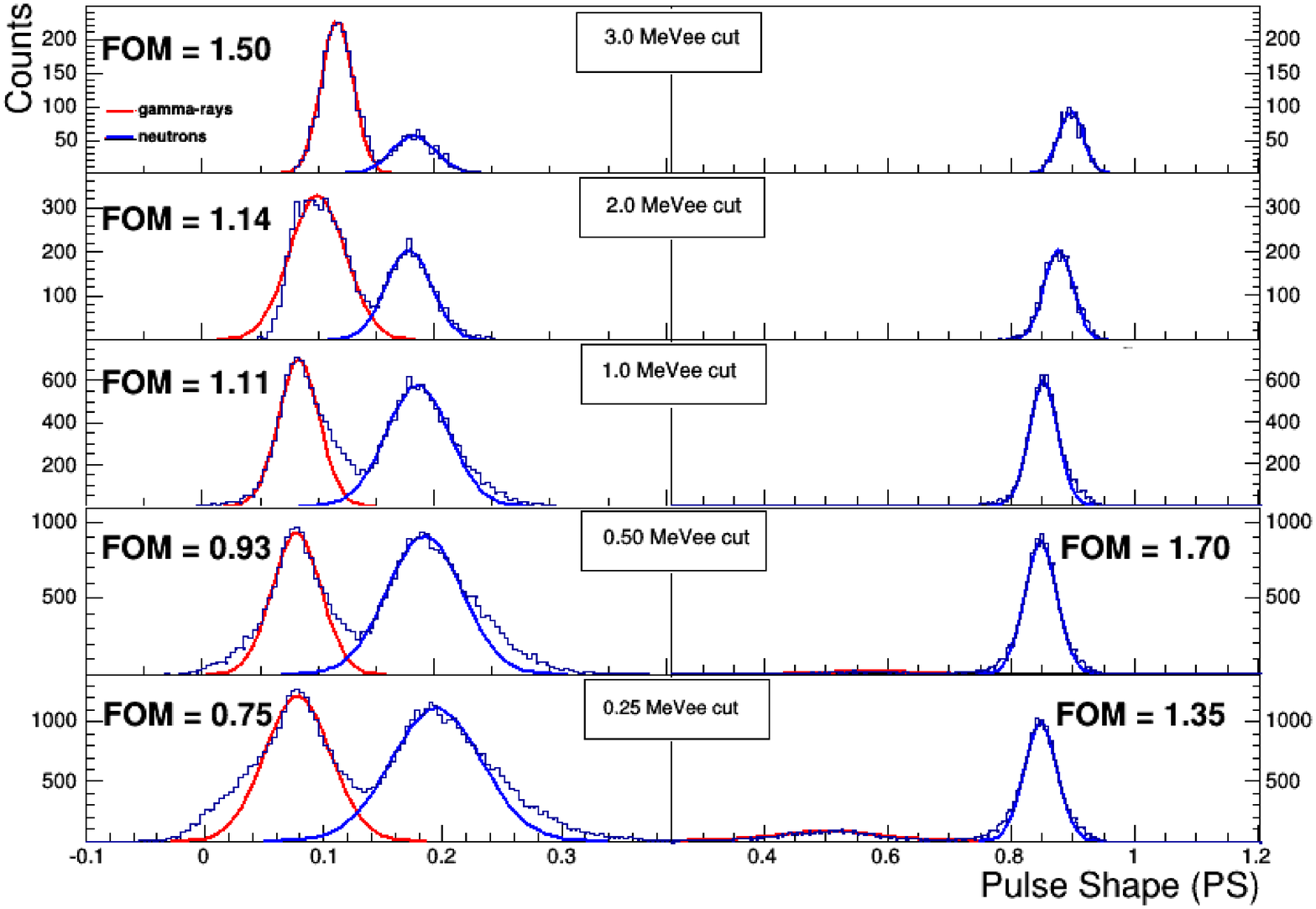}}
\caption{\label{figure:fom_projections}
A first comparison of the PSD achieved using the NE-213 reference cell (left 
column) and the $^{4}$He-based NDT (right column) via a FOM using the Pb-screened 
Am/Be source and varying the requirements on the amount of energy deposited in 
the detectors. Gamma-rays with energies up to 4.44~MeV deposit no more than
750~keV$_{ee}$ in the Arktis NDT which greatly facilitates fast-neutron 
particle identification.
}
\end{figure}

One-dimensional projections of PS have been obtained from the two-dimensional 
distributions shown in Fig.~\ref{figure:ndt_refcell_psd} for five different 
pulse-height thresholds. The resulting PS distributions for the NE-213 detector 
(left column) and Arktis NDT (right column) integrated from these thresholds 
are shown in Fig.~\ref{figure:fom_projections}. In each of the panels, wherever 
possible, two separate Gaussian functions have been fitted to the data -- one 
corresponding to gamma-rays (red) and one corresponding to neutrons (blue). A 
standard figure-of-merit (FOM) has been used to quantify the quality of the 
PSD as a function of deposited-energy cut. This FOM is given by the separation 
between the gamma-ray and neutron peaks divided by the sum of the FWHM of these 
peaks. 

As can be seen in the NE-213 data (left column), the FOM improves from 0.75 to 
1.5 as the requirement on the light yield increases from 0.25 to 3.0 MeV$_{ee}$. 
In comparison, the Arktis NDT FOM (right column) improves from 1.35 to 1.70 as 
the requirement on the light yield increases from 0.25 to 0.50 MeV$_{ee}$. Once 
the requirement on the amount of deposited energy exceeds 0.75 MeV$_{ee}$, the 
gamma-ray peak is no longer visible in the Arktis NDT, and the signal is 
unambiguously neutron. Again, the fact that gamma-rays with energies up to 
4.44~MeV deposit no more than 750~keV$_{ee}$ in the Arktis NDT is striking. 
This property of the detector greatly facilitates the identification of fast 
neutrons depositing more energy than this value.

\section{Summary}
\label{section:summary}

A first comparison between the PSD characteristics of a novel $^{4}$He-based 
high-pressure gas scintillation detector and a standard NE-213 
liquid-scintillator reference detector has been performed. A Pb-screened Am/Be 
mixed-field neutron and gamma-ray source was used to irradiate the detectors 
and a high-resolution scintillation-pulse digitizer was used to optimize the 
PSD using the tail-to-total method. The NE-213 liquid-scintillator reference 
cell was differentially very sensitive to the incident gamma-rays and 
registered a wide range of scintillation-light yields up to 4.4~MeV$_{ee}$.
In contrast, the $^{4}$He-based NDT was designed to have has a low gamma-ray 
sensitivity. 
It registered a maximum scintillation-light yield of 750~keV$_{ee}$ for the 
same distribution of incoming gamma-rays. The PSD obtained with the NE-213 
liquid-scintillator reference cell, facilitated by using the digitizer, was 
good. Clear separation between neutrons and gamma-rays was obtained down to 
about 0.5~MeV$_{ee}$. The PSD obtained with the $^{4}$He-based NDT was 
excellent. Clear separation between neutrons and gamma-rays was obtained down 
to 0.1~MeV$_{ee}$. Most striking was the fact that gamma-rays of energies up 
to 4.44~MeV resulted in scintillation-light yields of no more than 750~keV$_{ee}$ 
in the NDT. As a result, a simple threshold cut above 750~keV$_{ee}$ was 
sufficient to distinguish fast neutrons from gamma-rays in this region. For 
fast-neutron detection, $^{4}$He-based high-pressure gas scintillation detectors 
such as the NDT thus have a clear advantage over liquid-scintillator detectors 
such as the NE-213 reference cell when the scintillation-light yield is greater 
than 750~keV$_{ee}$.

The next step in our investigations shall involve an expanded systematic 
comparison of the PSD obtained with these two scintillators using a FOM as a 
function of scintillation-light yield. The scintillation-light yield as a 
function of neutron energy must also be established if the potential for 
spectroscopy is to be investigated.  In order to perform these investigations, 
knowledge of the incident neutron energies is required. We have recently 
successfully tested a technique for performing such irradiations with 
Be-compound neutron sources~\cite{scherzinger14} which relies on well-understood 
shielding, coincidence, and time-of-flight measurement techniques to produce 
a polychromatic energy-tagged neutron beam.

\section*{Acknowledgements}
\label{acknowledgements}

We thank the Photonuclear Group at the MAX IV Laboratory for providing access
to their experimental hall and Am/Be source. We acknowledge the support of the
UK Science and Technology Facilities Council (Grant nos. STFC 57071/1 and 
STFC 50727/1).

\bibliographystyle{elsarticle-num}

\end{document}